\documentclass[review, number]{elsarticle}

\journal{Physica D}
\usepackage{graphicx}
\usepackage{amssymb}
\usepackage{amsmath}
\usepackage{latexsym}
\biboptions{sort&compress}

\begin{document}

\begin{frontmatter}

\title{Scaling exponents and phase separation in a nonlinear network model inspired by the gravitational accretion}

\author{Aleksandar Bogojevi\'c}

\author{Antun Bala\v{z}\corref{cor}}
\ead{antun@ipb.ac.rs}

\author{Aleksandar Beli\'c}

\address{Scientific Computing Laboratory, Institute of Physics Belgrade, University of Belgrade\\
Pregrevica 118, 11080 Belgrade, Serbia, {\tt http://www.scl.rs/}}

\cortext[cor]{Corresponding author.}

\begin{abstract}
We study dynamics and scaling exponents in a nonlinear network model inspired by the formation of planetary systems. Dynamics of this model leads to phase separation to two types of condensate, light and heavy, distinguished by how they scale with mass. Light condensate distributions obey power laws given in terms of several identified scaling exponents that do not depend on initial conditions. The analyzed properties of heavy condensates have been found to be scale-free. Calculated mass distributions agree well with more complex models, and fit observations of both our own Solar System, and the best observed extra-solar planetary systems.
\end{abstract}


\end{frontmatter}

\section{Introduction}
The studies of nonlinear systems in physics rely on the well known and developed approaches, but have also greatly benefited from the development of new methods which offer new insights and prospectives for understanding of the behavior of such systems. The approach based on the study of complex networks represents one of seminal examples for such advancements \cite{dorogovtsev, albertrmp}. To mention just the hallmarks, such as the study of the preferential attachment by Yule \cite{yule} and Simon \cite{simon}, and of the World Wide Wed \cite{diameterwww} in the pivotal work by Albert, Jeong and Barab\' asi, this approach allowed investigations of the emergence of scaling behavior in complex systems \cite{emergence}, utilized the concept of scale-free networks \cite{scale-free}, enabled studies of dynamics and topologies of evolving networks \cite{albertprl1, albertprl2}, and provided important new methods for applications ranging to biophysics \cite{metabolic, protein, gene1, gene2, neural} and techno-social networks \cite{m1, m2, m3}.

The level of details used to describe nonlinear systems is necessarily limited, both in analytic and numerical approaches. However, if carefully crafted, even the simplified models can provide valuable information on the behavior and properties of such systems \cite{internet, traffic1, urban, air, traffic2}. The formation of planetary systems is a well-known nonlinear system exhibiting chaotic behavior, and its modeling has taken on increased relevance due to the abundance of new observational data on extra-solar planets \cite{Marcy98,Lissauer02,Santos05}. Recent models of planet formation have tended towards greater sophistication, incorporating many complex phenomena. Such models aim to provide detailed understanding of various stages of planet formation, such as initial collapse of the protostar, interaction of dust and gas in the young disk, non-gravitational accretion of dust into mountain-sized objects, the separate processes of creating terrestrial planets, cores of gas giants, and gas accretion onto cores, and finally the interaction between formed planets and the remaining disk material \cite{Wetherill96b,Kokubo98,Perryman00,Lissauer05}.  This ``divide and conquer'' approach has yielded significant results and represents the state-of-the-art in the field.

The current paper has the complementary goal of applying the standard method of effective model building used for studying nonlinear systems to the exciting field of planetary system formation. We do this by developing and numerically simulating a simplified toy-model of the gravitationally-dominated phase of planetary system formation. The model's simplicity makes it possible to span a mass scale from mountain-sized to planet-sized objects in a unique framework, to obtain detailed statistics by averaging over hundreds of runs, as well as to study how accretion outcomes depend on initial conditions. We use the presented model as a heuristic tool for getting a handle on understanding the dominant properties of accretion, and for studying the observed power laws, e.g. between spin angular momentum and mass, in the size distribution of smaller objects in the Solar system etc.

The nonlinear model presented seeks to emulate the true spirit of effective model building in two key aspects. First, unlike in earlier attempts \cite{Dole70,Isaacman77,Laskar00}, the employed interaction criterion is not given \emph{ad hoc} but rather follows from the underlying micro physics. Second, the key requirement put before the model is that it correctly reproduces the qualitative behavior and uncovers functional connections between relevant dynamical quantities. We use such qualitative understanding to distinguish different types of condensates appearing in accretion and to study the dependence of their properties on the initial conditions in the protoplanetary disk. The ultimate goal of this is to obtain an initial handle on the classification of possible types of planetary systems.

Before launching into details, we offer a brief preview of how our model matches observations. First, the spin angular momentum as a function of mass obeys a power law with a scaling exponent approximately matching that seen in the Solar System (Fig.~\ref{fig:spin}). Second, the ratios of the masses of the three heaviest planets in our Solar System can be reproduced using the model's single input parameter $K$ defined below (Fig.~\ref{fig:mass} and Table~\ref{tab:best-fits}). Third, using the obtained values of parameter $K$ we correctly recover (Table~\ref{tab:best-fits}) the accepted mass of the Minimum Mass Solar Nebula \cite{Weidenschilling05}. The analyzed toy-model also successfully fits the three heaviest masses of extra-solar systems for which several planets are known (Table~\ref{tab:best-fits}). Finally, the mass distribution of ``light'' condensates (Fig.~\ref{fig:delta}) exhibits a power law behavior in agreement with the observed size distribution of main-belt asteroids \cite{Ivezic01}. It is well known that most of the main-belt asteroids, except the largest ones, have been shattered after the formation of the main-belt and that their size distribution is dominated by fragmentation, not accretion. The above result suggests the plausible scenario in which the fragmentation process preserves the power law distribution in size.

We believe that the presented nonlinear network model and the obtained scaling laws and scaling exponents, as well as the observed phase separation could contribute to the understanding of the complex process of planetary formation and its main features.

\section{The model}
The presented model starts from a given planar distribution of $N$ initial particles of equal mass and with no spin. The initial particles have a uniform angular distribution, while the radial distribution is given by $\rho(r)$ which is normalized according to $\int_0^\infty dr\,\rho(r)=M_P$, where $M_P$ is the total mass of the protoplanetary material.

The $N$-body dynamics is simplified by dividing it into two pieces---free propagation on Keplerian trajectories and instantaneous binary mergers. For simplicity, we assume that all trajectories are nearly circular, and neglect their eccentricities. The binary merger proceeds if the two particles satisfy an interaction criterion given below. Although all orbits in the model are circular and do not cross, we assume that two bodies whose orbits are close enough (as will be defined by the interaction criterion) can merge, and that the growing protoplanets have some radial reach that allows them to accrete neighboring bodies. Another limitation of the model is that it assumes that there is no radial migration of the bodies, while, in fact, small bodies will drift radially due to nebula gas drag, and larger protoplanets will migrate inwards due to their gravitational interactions with the nebula.

The result of the merging of bodies with masses $m_1$ and $m_2$, at positions $r_1$ and $r_2$, and spins $s_1$ and $s_2$, is a new body with mass $m$, position $r$ and spin $s$. The properties of the new body follow from mass, energy and angular momentum conservation. Mass conservation gives $m=m_1+m_2$. Expressing angular momenta in units of $\sqrt{M_*G}$, where $M_*$ is the mass of the star, the orbital angular momentum of a body of mass $m$ at a distance $r$ from the star equals $\ell =m\sqrt{r}$. All bodies are assumed to have no initial spin, and after each merger the excess angular momentum $\Delta\ell=m_1\sqrt{r_1}+m_2\sqrt{r_2}-(m_1+m_2)\sqrt{r}$ is, according to the angular momentum conservation law, converted into the spin of the new body, $s=s_1+s_2+\Delta\ell$. The position $r$ follows from energy conservation,
\begin{equation}
-\frac{GM_* m_1}{2r_1}-\frac{GM_* m_2}{2r_2}=-\frac{GM_*(m_1+m_2)}{2r}+Q\, ,
\end{equation}
where $Q$ is the thermal energy corresponding to the heating of the merged body. We have neglected the much smaller contributions of the potential energy of the gravitational interaction of the condensing bodies, as well as kinetic energies due to spin. $Q\ge 0$ implies that the merging position satisfies $r\le r_\mathrm{s}$, where $m/r_\mathrm{s}=m_1/r_1+m_2/r_2$. In addition, $r_\mathrm{s}<r_0$, where $r_0$ is the merging position leading to zero spin, i.e. $m\sqrt{r_0}=m_1\sqrt{r_1}+m_2\sqrt{r_2}$. As a result, the thermodynamic requirement $Q\ge 0$ implies that, within our model, spin is necessarily positive. From now on we use the merging point $r=r_\mathrm{s}$, corresponding to $Q=0$.

The above relations completely specify the kinematics. The merging criterion encoding the dynamics of the model follows from what is an essentially dimensional analysis of Newton's laws, and is determined as follows: two bodies merge if $|F\triangle t|\gtrsim\,|\triangle p|$, where $F$ is the characteristic gravitational force between the bodies during interaction, $\triangle t$ the characteristic time for the merger, and $\triangle p$ the resulting change in momentum.  We assume that particles interact only at their closest approach, and so disregard all dependence on orbital position and find $|F|\sim Gm_1m_2/(r_1-r_2)^2$. The characteristic time is $\triangle t \sim |\triangle r|/|\triangle v|$, where $|\triangle r|=|r_1-r_2|$, and $|\triangle v|=\sqrt{GM_*}\,|1/\sqrt{r_1}-1/\sqrt{r_2}|$. Similarly, the change in momentum due to merger equals $|\triangle p|=\sqrt{GM_*}\,|(m_1+m_2)/\sqrt{r}-m_1/\sqrt{r_1}-m_2/\sqrt{r_2}|$. The complete merging criterion becomes
\begin{equation}
\label{interaction}
\frac{|r_1-r_2|}{m_1 m_2}
\left|\frac{1}{\sqrt{r_1}}-\frac{1}{\sqrt{r_2}}\right|
\left|\frac{m}{\sqrt{r_s}}-\frac{m_1}{\sqrt{r_1}}-\frac{m_2}{\sqrt{r_2}}\right|\le K\, .
\end{equation}
Merging stops when no two particles satisfy Eq.~(\ref{interaction}). Dynamics within our effective gravitational accretion model
is driven by a single parameter $K$. If we express masses $m_1$ and $m_2$ in units of $M_P$, the parameter $K$ becomes dimensionless, and is proportional to $M_P/M_*$. Since our merging process follows from a dimensional analysis, the proportionality factor between $K$ and $M_P/M_*$ cannot
easily be determined, but is assumed to be close to unity. As already pointed out, the model neglects radial migration, which is quite sensitive to masses of bodies.
This will promote differential migration, and the added mobility will enhance the bodies' ability to accrete. Therefore, radial migration would render $K$ parameter to be mass-dependent, and the system could not be described via a single value of $K$. However, these effects are neglected in the simplified model employed here.

The interaction criterion is homogeneous with respect to changes of both mass and distance scales. We fix the mass scale by setting $M_P=1$, i.e. by expressing the masses of all the bodies in units of $M_P$. Distance scales are fixed by our choice of $\rho(r)$. For $m_1\gg m_2$, the derived criterion implies that $m_1$ merges with any body within the $r_1\pm\delta(r_1,m_1)$ feeding zone, the Hill's radius, where $\delta(r,m)= 2rK^{1/4}m^{1/4}\sim r(m/M_*)^{1/4}$. The same approximative result was also obtained in some early studies \cite{Dole70,Dole61}, while the correct scaling for the Hill's radius $\delta(r,m)\sim r(m/M_*)^{1/3}$ follows from the exact treatment of the restricted three-body problem \cite{Lissauer1993}. This is reasonably close to the result obtained from our simplified model, and indicates that its single interaction criterion, to some extent, effectively encodes some of major ingredients in gravitational accretion, such as particle collisions and gas capture. We stress that the toy model does not include important effects due to resonances and tidal lock.

\section{The algorithm and implementation details}
The straightforward way to simulate the presented model
\cite{b1, b2, b3} would be to generate the
positions of all $N$ initial particles according to initial mass
distribution $\rho(r)$ and then to randomly pick the pairs and
merge them if the interaction criterion is satisfied. The random
number generator used in this work is RAN3 described in Press,
{\it et al.} \cite{nr}. This process would continue until no further
merging was possible, i.e. no pair of bodies satisfied the
criterion. Such a strategy (definition algorithm) has been
investigated and the memory requirements needed to simulate $N$
body accretion scale as $O(N)$, whereas computing time has been
found to be of order $O(N^{2.2})$. This makes it impractical for
the study of sufficiently large systems.

Fortunately, there is a more efficient way to simulate our model.
It is not necessary to specify all the $N$ bodies at the very
beginning, due to the fact that we are dealing with a two body
interaction criterion. For example, in the very first merger the
position of the $N-2$ spectators are irrelevant, and need not be
generated at that time. After the question of this merging is
resolved another particle is added and so on. At each step the
possible merging of the newly added particle with the ones already
present is investigated. This operation is local, i.e. if the
newly introduced particle merges that can only happen with one of
the two particles with the nearest radial distance, and for this
reason it is useful to keep particles sorted according to
increasing $r$ throughout the simulation. Obviously, the positions
and masses of particles change after merging, making further
merging possible. After all the merging possibilities are
exhausted a new particle is added to the system, and the procedure
is recursively repeated until all $N$ initial particles are
considered. Although we can't prove the strict equivalence of
these two simulation schemes numerical evidence shows that the
results are equivalent within statistical errors.

%------------------------------------------------------------------------------
\begin{figure}[!t]
\centering
\includegraphics[width=8cm]{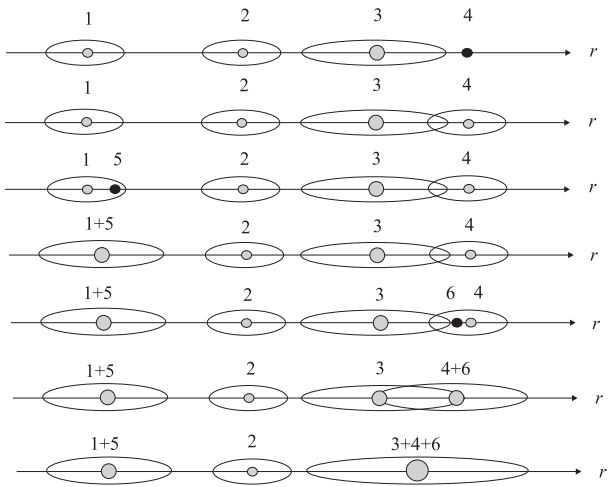}
\caption{Schematic presentation of the order of introduction of initial particles used in the employed algorithm.}
\label{fig:order}
\end{figure}
%------------------------------------------------------------------------------

Fig.~\ref{fig:order} gives a schematic presentation of how initial
particles are introduced. Black dots represent newly introduced
particles, gray dots represent existing particles, while ellipses
correspond to their regions of attraction for the capture of
initial particles. At the top line we introduce a new particle
that does not merge with the rest of the particles. The resulting
situation is shown in line two. Lines three and four depict a
typical merging. Lines five to seven show the introduction of a
new particle that leads to a two step merging cascade. In line two
we see that particles 3 and 4, while not interacting, have
overlapping regions of attraction for the capture of initial
particles. In cases like this we need to specify whether the
merging proceeds to the left or right. We have investigated both
the cases when all such merging is to the left and to the right.
The difference is quite small and may be absorbed into a change of
$K$. Throughout this paper we resolve the case of overlapping
regions of attraction by always merging to the left, i.e. to
smaller values of $r$.

%------------------------------------------------------------------------------
\begin{figure}[!b]
\centering
\includegraphics[width=8cm]{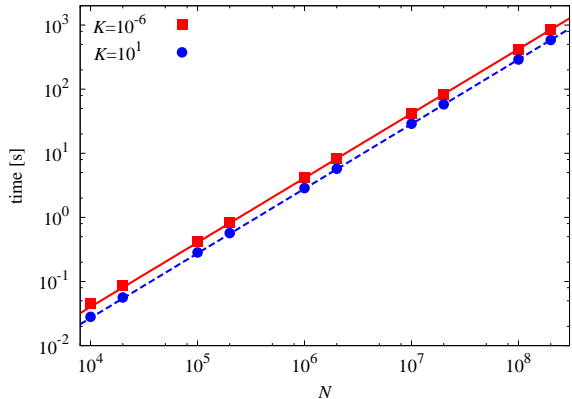}
\caption{The computing time (on a moderate Intel Xeon E5345 2.33 GHz processor) for a single run as a function of $N$ for $K=10^{-6}$ (top) and $K=10^1$ (bottom line). The two lines represent fits to the scaling law from Eq.~(\ref{eq:scaling}), with $a=3.8\cdot 10^{-6}$, $b=2.2\cdot 10^{-8}$ for $K=10^{-6}$ (full line) and
$a=2.5\cdot 10^{-6}$, $b=2.2\cdot 10^{-8}$ for $K=10^{1}$ (dashed line).}
\label{fig:time}
\end{figure}
%------------------------------------------------------------------------------

The memory required for this algorithm is of the order $O(n)$, where
$n$ (typically $n\ll N$) is the final number of condensates. On
the other hand, we expect the computing time to be
\begin{equation}
\label{eq:scaling}
t=N(a+b\log N)\, .
\end{equation}
The overall factor of $N$ comes from the loop over
$N$ initial particles. The term in brackets represents the time
for the calculations inside the loop, i.e. for a single particle.
The $b\log N$ term comes from keeping particles sorted according
to increasing $r$ throughout the simulation, while the constant
term $a$ represents the number of other operations inside the
loop, regardless of sorting. Obviously, for $N\gg 10^{a/b}$ the
term containing the logarithm will dominate and the algorithm will
be $O(N\log N)$. However, since in our code $a/b> 100$ that
regime is never reached in practice and, for the considered
numbers $N=10^3-10^{10}$, the algorithm is $O(N)$, as can be seen
from Fig.~\ref{fig:time}.

\section{Numerical results}

%------------------------------------------------------------------------------
\begin{figure}[!b]
\centering
\includegraphics[width=6cm]{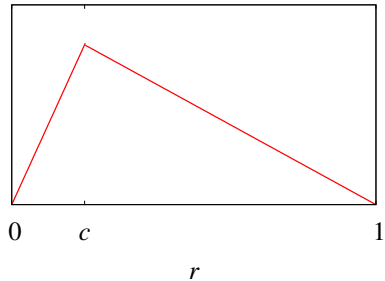}
\caption{Illustration of the triangular-shaped distributions on $r\in [0,1]$ peaked at $c$.}
\label{fig:triangle}
\end{figure}
%------------------------------------------------------------------------------

Our model depends on the parameter $K$, the number of initial bodies $N$, and the initial mass distribution $\rho$. In this letter we investigate the robustness of gravitational accretion on the choice of $\rho$. We shall show that several important consequences of accretion are independent of, or depend very weakly on, the specific form of $\rho$. This is particularly important since very little is known about the true conditions at the start of accretion. To demonstrate this we conducted numerical simulations using our SOLAR code \cite{solar} on a wide range of initial mass distributions (Table~\ref{tab:massexp}): triangular-shaped distributions on $r\in [0,1]$ peaked at $c=10^{-1}$, $10^{-2}$, $10^{-3}$ (see Fig.~\ref{fig:triangle}); uniform distribution on $r\in [0,1]$; $\rho(r)\propto r/(1+r^4)$; $\rho(r)\propto r(2-r)$ (for $r\in [0,2]$). The listed distributions have not been chosen for physical reasons, but so as to investigate the dependence of outcomes on drastic changes of initial conditions. All the presented graphs of numerical simulations were done on the triangular-shaped distribution peaked at $c=10^{-1}$ which we denote `triangle 1', although our results do not change much if other distributions are used.

%------------------------------------------------------------------------------
\begin{figure}[!b]
\centering
\includegraphics[width=7.7cm]{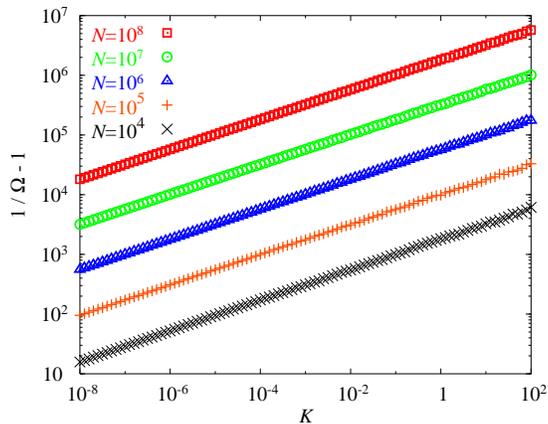}
\caption{$1/\,\Omega -1$ as a function of $K$ for $N$ from $10^4$ to $10^8$ (average over 100 runs, error bars shown).}
\label{fig:omega}
\end{figure}
%------------------------------------------------------------------------------

We start by identifying a set of quantities that characterize the final distribution of condensates. The first such quantity is the ratio of the number of final and initial bodies $\Omega = n/N$. From Fig.~\ref{fig:omega} we see that $1/\,\Omega -1$ is given by a power law in both $N$ and $K$. The fit gives
\begin{equation}
1/\Omega-1\propto N^\alpha K^\beta\, .
\label{omega}
\end{equation}
The `triangle 1' distribution gives $\alpha=0.744(1)$, and $\beta=0.251(1)$. From Table~\ref{tab:massexp} we see that the scaling exponents depend very weakly on $\rho$. $\Omega$ decreases monotonically with $K$ from its maximal value of 1 at $K = 0$, to the minimum $1/N \sim 0$ at large $K$, the two extremes corresponding respectively to no accretion, and the collapse of all the material into a single body. Thus the model's single input parameter $K\sim M_P/M_*$ regulates the amount of accretion taking place.

The power-law behavior of the quantity $\Omega$ comes about through a runaway growth process \cite{Kokubo96}. Similar mechanisms have been observed in completely different realms, for example when a graph (of the node-edge variety) grows via preferential attachment of new nodes to existing nodes of greater weight. Such structure occurs in systems such as power grids, the Internet, and the WWW \cite{diameterwww, emergence, scale-free}.

\begin{table}[!t]
\caption{Mass distribution scaling exponents for different initial densities: triangles on $r\in [0,1]$ peaked at $c=10^{-1}$, $10^{-2}$, $10^{-3}$; uniform on $r\in [0,1]$; proportional to $r/(1+r^4)$; proportional to $r(2-r)$. Numbers in brackets denote statistical errors on last digit.}
\label{tab:massexp}
\begin{center}
\begin{tabular}{l|l|l|l}
\hline
\multicolumn{1}{c|}{$\rho(r)$}&\multicolumn{1}{c|}{$\alpha$}& \multicolumn{1}{c|}{$\beta$}&\multicolumn{1}{c}{$\tau$}\\
\hline
triangle 1 &$0.744(1)$ &$0.251(1)$ &$1.22(1)$ \\
triangle 2 &$0.758(2)$ &$0.256(3)$ &$1.20(1)$ \\
triangle 3 &$0.775(2)$ &$0.264(6)$ &$1.19(1)$ \\
uniform &$0.751(1)$ &$0.247(1)$ &$1.21(1)$ \\
$r/(1+r^4)$&$0.735(8)$ &$0.243(3)$ &$1.22(1)$ \\
$r(2-r)$ &$0.749(2)$ &$0.249(1)$ &$1.21(1)$ \\
\hline
\end{tabular}
\end{center}
\end{table}

%------------------------------------------------------------------------------
\begin{figure}[!t]
\centering
\includegraphics[width=7.7cm]{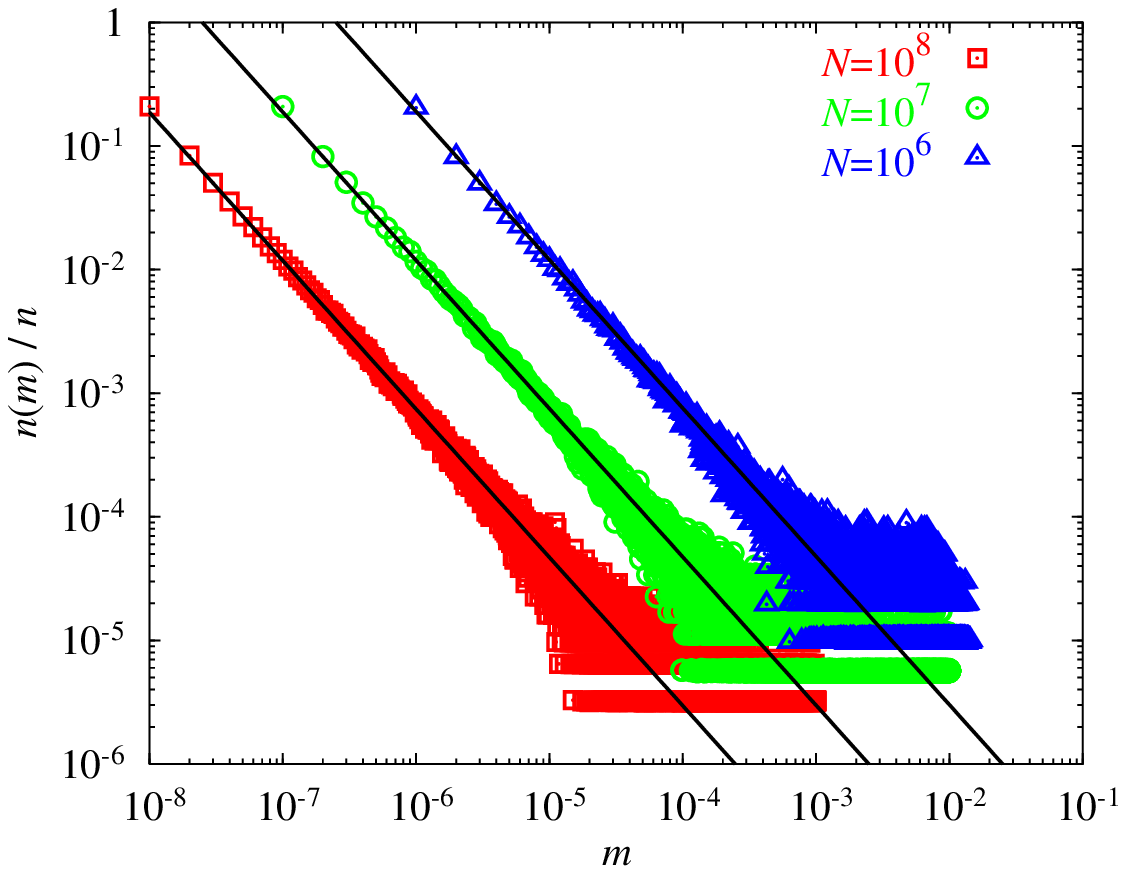}
\includegraphics[width=7.7cm]{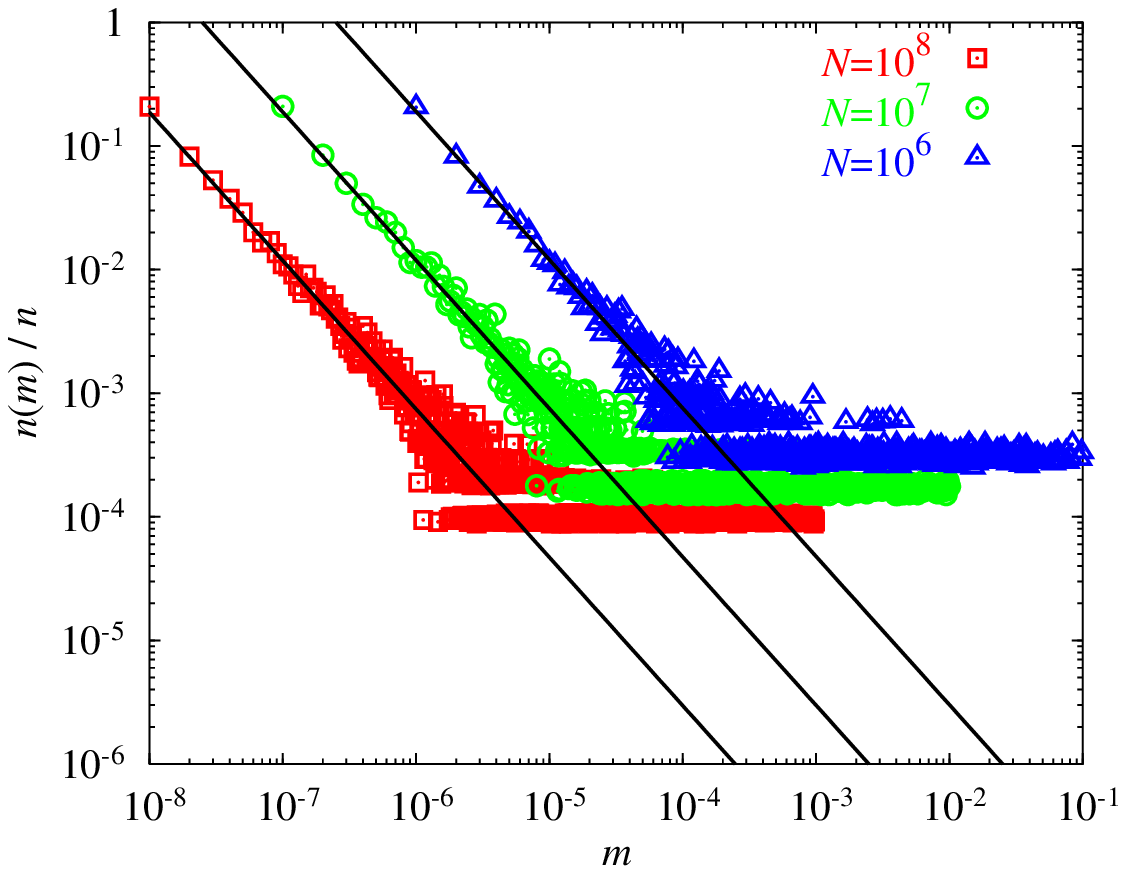}
\caption{Relative number of condensates as a function of mass for $N$ from $10^6$ to $10^8$ (average over 100 runs). Top: Weak condensation ($K = 10^{-7}$), almost all condensates belong to the light class and exhibit scaling. Bottom: Strong condensation ($K = 0.04$), leading to Solar-like planetary systems with both light and heavy condensates.}
\label{fig:delta}
\end{figure}
%------------------------------------------------------------------------------

The quantity $\Omega$ is a global property of condensates. A more detailed understanding of their structure is achieved by studying their distribution in mass. From Fig.~\ref{fig:delta} we see that the mass distribution of light condensates fits well to the $K$-independent power law
\begin{equation}
n(m_i)/n \propto (Nm_i)^{-\tau}\, .
\label{tau}
\end{equation}
Condensates that scale according to this law are designated as light, those that do not as heavy, the dividing line being at critical mass $m_c$. For the `triangle 1' distribution $\tau=1.22(1)$, corresponding to a size distribution with exponent 3.66 (since $m\sim d^3$), in good agreement both with the observed size distribution of main belt asteroids \cite{Ivezic01}, and with more detailed models of planet accretion that predict a {\em cumulative} distribution in mass with exponent $2.5\pm 0.4$ \cite{Kokubo96}. Furthermore, from Table~\ref{tab:massexp} we see that this is another example of an exponent essentially independent of initial conditions.

To leading order, the average mass of light condensate does not depend on $m_c$ and is approximately equal to $\tau/(\tau-1)N$. The spread of the mass of light condensates about that mean depends on $m_c$, vanishing with it as $(m_c)^{1-\tau/2}$. The total mass of the light condensates is given by $M_\mathrm{light}\approx \overline{m}_\mathrm{light}n_\mathrm{light}\approx \frac{\tau}{\tau-1}\,\Omega$. As we can see from Table~\ref{tab:massexp}, $\alpha\in(0,1)$ and it follows that in the large $N$ limit, the number of light condensates goes to infinity while their contribution to the total mass vanishes.

%------------------------------------------------------------------------------
\begin{figure}[!t]
\centering
\includegraphics[width=7.7cm]{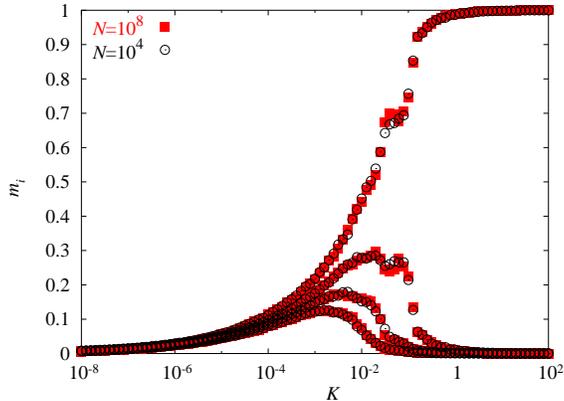}
\caption{Masses of four heaviest condensates $m_i$ as functions of $K$ for $N=10^4$ and $10^8$ (average over 100 runs).}
\label{fig:mass}
\end{figure}
%------------------------------------------------------------------------------

Formed planetary systems are dominated by a relatively small number of extremely heavy condensates, which we designate as planets. In our effective model, we investigated their mass, position and spin, using runs with up to $N=10^{10}$ particles. Fig.~\ref{fig:mass} shows the masses of the four heaviest condensates as a function of the condensation parameter $K$. As can be seen, the masses of the heaviest planets are essentially independent of $N$ for $N\gtrsim 10^4$, i.e. they have already converged to the continuum limit.

For $K=0.04$ the simplified model shows good agreement with the observed ratios of the masses of Jupiter, Saturn, Neptune, and even Uranus (in order of mass). Note that $K=0.04 \sim M_P/M_*$ is consistent with mass estimates of the Minimum Mass Solar Nebula \cite{Weidenschilling05}. Table~\ref{tab:best-fits} demonstrates that a similar fitting can be made to two currently observed extra-solar systems for which several planets are known, namely HD 160691 and 55~Cnc.

\begin{table}[!t]
\caption{Fitting $K$ from Fig.~\ref{fig:mass} to the planetary mass ratios (sorted by mass) of recently discovered extrasolar systems with several observed planets. For Sun, $K=0.04$ corresponds to the accepted value of the Minimum Mass Solar Nebula.}
\label{tab:best-fits}
\begin{center}
\begin{tabular}{l|l|l|l}
\hline
\multicolumn{1}{c|}{Name}&\multicolumn{1}{c|}{$m_1$}&\multicolumn{1}{c|}{$m_2$}&\multicolumn{1}{c}{$m_3$}\\
\hline
$K$=0.025&0.58(1)&0.276(7)&0.099(6)\\
HD 160691&0.581&0.313&0.098\\
\hline
$K$=0.04&0.699(9)&0.238(7)&0.040(3)\\
Sun&0.712&0.213&0.038\\
\hline
$K$=0.11&0.80(2)&0.18(2)&0.020(1)\\
55 Cnc&0.789&0.158&0.044\\
\hline
\end{tabular}
\end{center}
\end{table}
%------------------------------------------------------------------------------
\begin{figure}[!t]
\centering
\includegraphics[width=7.7cm]{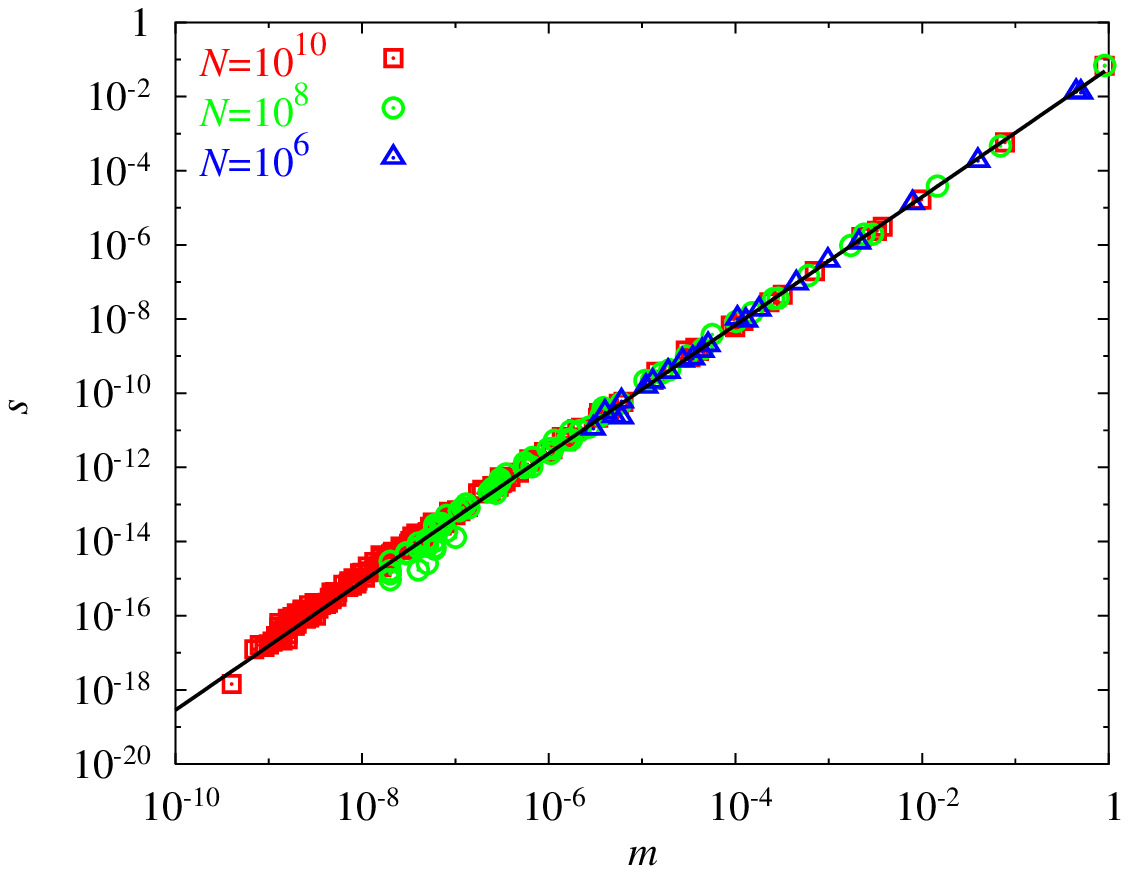}
\includegraphics[width=7.7cm]{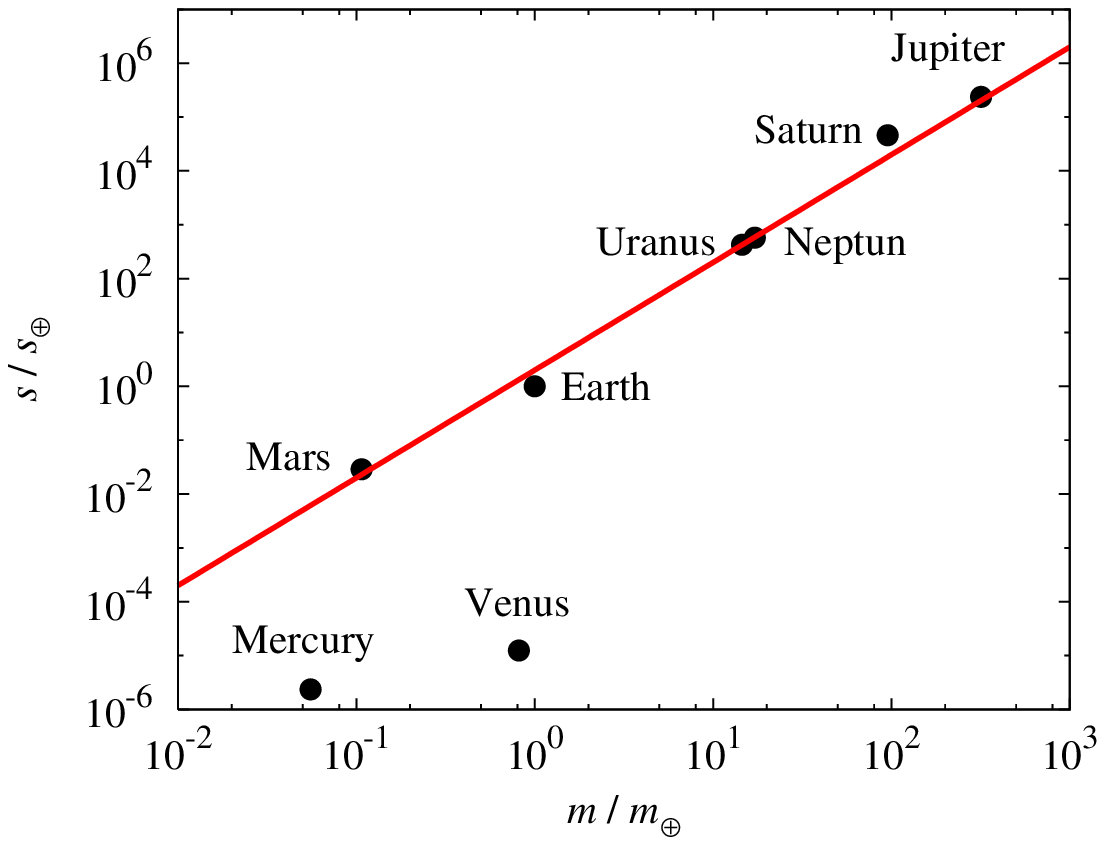}
\caption{Top: Spin of condensates as a function of mass for $K=0.04$ and $N=10^6,\, 10^8,\, 10^{10}$. The data fit to $s\propto m^{1.72}$. Bottom: Spin vs.~mass in the Solar system, expressed in relative units of Earth's mass $m_\oplus$ and spin $s_\oplus$. The planets in the Solar system fit to $s\propto m^2$.}
\label{fig:spin}
\end{figure}
%------------------------------------------------------------------------------

We stress that $K$ is the only parameter in the model, and that by fixing it, all the other results become predictions. Choosing $K$ in the same way for different $\rho$'s leads to roughly the same planetary masses. The weak dependence of the planetary masses on the initial mass distribution is the reason why we get good agreement with Solar system data even for physically unrealistic distributions such as `triangle 1'. However, we stress again that the radial migration of bodies is neglected in our model, and therefore good agreement with Solar system data is somewhat fortunate.

We next focus on the spin. Both the dynamics and initial conditions of our model are planar, hence, the planets can only spin up or down. Recall that in our effective model, thermodynamics constrains spins to be positive. The top plot in Fig.~\ref{fig:spin} displays the spin of the condensates as a function of their mass, illustrating that spin is dependent on $N$ only in that larger $N$ allows the existence of smaller objects.  The data fits to
\begin{equation}
s\propto K^\epsilon m^\omega\, .
\end{equation}

The `triangle 1' distribution gives $\omega=1.72(2)$ and $\epsilon=0.45(1)$. From Table~\ref{tab:spinexp} we see that $\epsilon$ is another example of a scaling exponent independent of initial distribution, while $\omega$ displays very weak dependence varying from 1.7 to 1.9 for the wide class of initial distributions considered. Note also that the spin of a condensate does not depend on the condensate's location, but only on its mass. The bottom plot in Fig.~\ref{fig:spin} shows that the planets in the Solar system obey the same kind of spin-mass dependence with exponent $\omega=2$ \cite{Brosche63}. Mercury does not satisfy the above spin-mass relation due to tidal lock effects. As we have already noted, our simplified model does not take into consideration tidal forces. Venus, on the other hand, is thought to have suffered a single large, spin-changing collision during the late stages of its formation \cite{Dormand87}. Such collisions are extremely rare and could not be seen in model predictions averaged over 100 runs. Note that Pluto is now considered a `dwarf planet' and not expected to obey the same scaling law, so we did not include it in Fig.~\ref{fig:spin}. However, although the agreement with Solar system data is quite good, we have to stress that a large number of effects was neglected in the model, such as spin-orbit resonances, tidal interactions with satellites, stochastic giant impacts, viscous accretion of circumplanetary gas disks, and these unmodeled effects would probably change the outcome.

\begin{table}[!t]
\caption{Spin distribution scaling exponents for different initial densities: triangles on $r\in [0,1]$ peaked at $c=10^{-1}$, $10^{-2}$, $10^{-3}$; uniform on $r\in [0,1]$; proportional to $r/(1+r^4)$; proportional to $r(2-r)$. Numbers in brackets denote statistical errors on last digit.}
\label{tab:spinexp}
\begin{center}
\begin{tabular}{l|c|c}
\hline
\multicolumn{1}{c|}{$\rho(r)$}&\multicolumn{1}{c|}{$\epsilon$}&\multicolumn{1}{c}{$\omega$}\\
\hline
triangle 1 &$\ 0.45(1)\ $ &$\ 1.72(2)\ $\\
triangle 2 &$0.44(1)$ &$1.78(2)$\\
triangle 3 &$0.44(1)$ &$1.90(2)$\\
uniform &$0.39(1)$ &$1.90(1)$\\
$r/(1+r^4)\quad$&$0.43(1)$ &$1.73(3)$\\
$r(2-r)$ &$0.44(1)$ &$1.71(2)$\\
\hline
\end{tabular}
\end{center}
\end{table}

Historically, radial distributions such as Bode's law have played a large role in describing the Solar System. Our model shows, however, that unlike mass distributions, radial distributions depend very strongly on initial conditions. In fact, this could be used in the future to obtain information about the `true' initial conditions. The strong dependence on position probably reflects planetary migration that has been seen in more detailed hydrodynamic models as following from planet disk interactions, and that have been used to explain the existence of so-called `hot Jupiter' extra-Solar planets. Within our model, migration is achieved through a cascade of mergers each of which leads to a slight change in a planetesimal's position.

\section{Conclusions}
We have presented and analyzed a simple, one-parameter network-based model of planetary system formation through gravitational accretion. Analytical comparisons with the restricted three-body problem and an analysis of the model's outcomes suggest that the toy model captures the main features of gravitational accretion. The model's simplicity allows for efficient implementation, large numbers of initial particles, and the study of a wide range of initial conditions, thus making possible the analysis of dominant properties common to planetary formation. The presented model leads to phase separation to two distinct types of condensates which appear dynamically, and which are distinguished by how they scale with the number of initial particles $N$. Several important properties of both light and heavy condensates have been analyzed. The scaling exponents and functional relations between dynamical variable uncovered have been shown to be in good agreement with observations. An important property of the model is the weak dependence of scaling exponents on the initial mass distribution. This is particularly important because of our limited knowledge on initial conditions at the start of gravitational accretion.

\section*{Acknowledgments}
The authors acknowledge useful discussions with Dr.~Wayne Hayes from the Department of Mathematics at Imperial College London. This work was supported in part by the Ministry of Education, Science, and Technological Development of the Republic of Serbia, under project No.~ON171017. Numerical simulations were run on the AEGIS e-Infrastructure, supported in part by European Commission through FP7 projects PRACE-1IP, PRACE-2IP, PRACE-3IP, HP-SEE, and EGI-InSPIRE.


\begin{thebibliography}{19}

\bibitem{dorogovtsev}
S.~N. Dorogovtsev and A.~V. Goltsev,
Rev. Mod. Phys. {\bf 80}, 1275 (2008).

\bibitem{albertrmp}
R. Albert and A.-L. Barab\' asi,
Rev. Mod. Phys. {\bf 74}, 47 (2002).

\bibitem{yule}
G.~U. Yule, Philos. Trans. R. Soc. London, Ser.~B {\bf 213}, 21 (1925).

\bibitem{simon}
H.~A. Simon, Biometrika {\bf 42}, 425 (1955).

\bibitem{diameterwww}
R. Albert, H. Jeong, and A.-L. Barab\' asi,
Nature {\bf 401}, 130 (1999).

\bibitem{emergence}
A.-L. Barab\' asi and R. Albert,
Science {\bf 286}, 509 (1999).

\bibitem{scale-free}
A.-L. Barab\' asi, R. Albert, and H. Jeong,
Physica A {\bf 272}, 173 (1999).

\bibitem{albertprl1}
R. Albert and A.-L. Barab\' asi,
Phys. Rev. Lett. {\bf 84}, 5660 (2000).

\bibitem{albertprl2}
R. Albert and A.-L. Barab\' asi,
Phys. Rev. Lett. {\bf 85}, 5234 (2000).

\bibitem{metabolic}
H. Jeong, B. Tombor, R. Albert, Z.~N. Oltvai, and A.-L. Barab\' asi,
Nature {\bf 407}, 651 (2000).

\bibitem{protein}
H. Jeong, S. Mason, A.-L. Barab\' asi, and Z.~N. Oltvai,
Nature {\bf 411}, 41 (2001).

\bibitem{gene1}
T. Ochiai, J.~C. Nacher, and T. Akutsu,
Phys. Lett. A {\bf 330}, 313 (2004).

\bibitem{gene2}
T. Ochiai, J.~C. Nacher, and T. Akutsu,
Phys. Lett. A {\bf 339}, 1 (2005).

\bibitem{neural}
J. Lu, D.~W.~C. Ho, and M. Liu,
Phys. Lett. A {\bf 369}, 444 (2007).

\bibitem{m1}
M. Mitrovi\' c, G. Paltoglou, and B. Tadi\' c, Eur. Phys. J. B {\bf 77}, 597 (2010).

\bibitem{m2}
M. Mitrovi\' c, G. Paltoglou, and B. Tadi\' c, J. Stat. Mech. Theory Exp. P02005 (2011).

\bibitem{m3}
M. Mitrovi\' c and B. Tadi\' c, Physica A {\bf 391}, 21 (2012).

\bibitem{internet}
M.~\' A. Serrano, M. Bogu\~ n\' a, and A. D'az-Guilera,
Phys. Rev. Lett. {\bf 94}, 038701 (2005).

\bibitem{traffic1}
S.-M. Cai, G. Yan, T. Zhou, P.-L. Zhou, Z.-Q. Fu, and B.-H. Wang,
Phys. Lett. A {\bf 366}, 14 (2007).

\bibitem{urban}
M. Barth\' elemy and A. Flammini,
Phys. Rev. Lett. {\bf 100}, 138702 (2008).

\bibitem{air}
L. Lacasa, M. Cea, and M. Zanin,
Physica A {\bf 388}, 3948 (2009).

\bibitem{traffic2}
A.~K. Nandi, K. Bhattacharya, and S.~S. Manna,
Physica A {\bf 388}, 3651 (2009).

\bibitem{Marcy98}
G.~W. Marcy and R.~P. Butler,
Ann. Rev. Astron. Astrophys. {\bf 36}, 57 (1998).

\bibitem{Lissauer02}
J.~J. Lissauer, Nature {\bf 419}, 355 (2002).

\bibitem{Santos05}
N.~C. Santos, W. Benz, and M. Mayor,
Science {\bf 310}, 251 (2005).

\bibitem{Wetherill96b}
G.~W. Wetherill,
Astrophys. Space Sci. {\bf 241}, 25 (1996).

\bibitem{Kokubo98}
E. Kokubo and S. Ida,
Icarus {\bf 131}, 171 (1998).

\bibitem{Perryman00}
M.~A.~C. Perryman.
Rep. Prog. Phys. {\bf 63}, 1209 (2000).

\bibitem{Lissauer05}
J.~J. Lissauer,
Space Sci. Rev. {\bf 116}, 11 (2005).

\bibitem{Dole70}
S.~H. Dole,
Icarus {\bf 13}, 494 (1970).

\bibitem{Isaacman77}
R. Isaacman and C. Sagan,
Icarus {\bf 31}, 510 (1977).

\bibitem{Laskar00}
J. Laskar,
Phys. Rev. Lett. {\bf 84}, 3240 (2000).

\bibitem{Weidenschilling05}
S.~J Weidenschilling,
Space Sci. Rev. {\bf 116}, 53 (2005).

\bibitem{Ivezic01}
\v Z. Ivezi\' c, S. Tabachnik, R. Rafikov, etal.,
Astron. J. {\bf 122}, 2749 (2001).

\bibitem{Dole61}
S.~H. Dole,
ARS J. {\bf 31}, 214 (1961).

\bibitem{Lissauer1993}
J.~J. Lissauer,
Ann. Rev. Astron. Astrophys. {\bf 31}, 129 (1993).

\bibitem{b1}
A. Bala\v z, A. Beli\' c, A. Bogojevi\' c,
Publ. Astron. Obs. Belgrade {\bf 65}, 17 (1999).

\bibitem{b2}
A. Bala\v z, A. Beli\' c, A. Bogojevi\' c,
Publ. Astron. Obs. Belgrade {\bf 65}, 23 (1999).

\bibitem{b3}
A. Bala\v z, A. Beli\' c, A. Bogojevi\' c,
Publ. Astron. Obs. Belgrade {\bf 65}, 27 (1999).

\bibitem{nr}
W.~H. Press, S.~A. Teukolsky, W.~T. Vetterling, and B.~P. Flannery,
{\it Numerical Recipes: The Art of Scientific Computing}, 2nd edn, Cambridge University Press, New York (2007).

\bibitem{solar}
SOLAR code,
{\tt http://www.scl.rs/solar/}

\bibitem{Kokubo96}
E. Kokubo and S. Ida,
Icarus {\bf 123}, 180 (1996).

\bibitem{Brosche63}
P. Brosche,
Z. Astrophys. {\bf 57}, 143 (1963).

\bibitem{Dormand87}
J.~R. Dormand and J. McCue,
J. Brit. Astr. Assoc. {\bf 98}, 23 (1987).

\end{thebibliography}
\end{document}